\documentclass[submission,copyright,creativecommons]{eptcs}

\title{Overlapping Coalition Formation via Probabilistic Topic Modeling}
\author{Michalis Mamakos
\institute{Northwestern University\\ Evanston, IL, USA}
\email{mamakos@u.northwestern.edu}
\and
Georgios Chalkiadakis
\institute{Technical University of Crete\\
Chania, Greece}
\email{gehalk@intelligence.tuc.gr}
}

\def\titlerunning{Overlapping Coalition Formation via Probabilistic Topic Modeling}
\def\authorrunning{M. Mamakos \& G. Chalkiadakis}

\usepackage{times}

\usepackage{url}
\usepackage{graphicx}
\usepackage{amssymb}
\usepackage{bm}
\usepackage{amsmath}
\usepackage{amsthm}
\usepackage[linesnumbered, ruled]{algorithm2e}
\usepackage{mathtools}
\usepackage{float}
\usepackage{subfig}
\usepackage{changepage}
\usepackage{balance}

\newcommand{\real}{\mathbb{R}}
\newcommand{\natur}{\mathbb{N}}
\newcommand{\vecr}{\mbox{\boldmath $r$}}

\DeclarePairedDelimiter\floor{\lfloor}{\rfloor}

\theoremstyle{definition}
\newtheorem{example}{Example}

\begin{document}

\maketitle

\begin{abstract}
Research in cooperative games often assumes
that agents know the coalitional
 values with certainty, and that they can belong to one coalition only.
By contrast, this work assumes that
the value of a coalition is based on 
an underlying collaboration \emph{structure}
emerging due to existing but \emph{unknown}
\emph{relations} among the agents; and that
agents can 
form \emph{overlapping coalitions}.
Specifically, we first propose \emph{Relational Rules},
 a novel representation scheme 
for cooperative games with overlapping coalitions,
which encodes the aforementioned relations,
and which extends the well-known MC-nets representation
 to this setting.
We then present a novel decision-making method
for decentralized overlapping  coalition formation,
 which exploits \emph{probabilistic topic modeling}---and,
   in particular,
\emph{online Latent Dirichlet Allocation}.
By interpreting formed coalitions as documents,
 agents can effectively
 learn topics that correspond
to profitable collaboration structures.
 

\end{abstract}

\section{Introduction}

Cooperative game theory~\cite{chalkiadakis2011} 
provides a rich framework for the coordination 
of the actions of
self-interested
agents.
Despite the maturity of the related literature, it is usually assumed that
an agent can be a member of exactly one coalition.
Nevertheless, in many real-world scenarios this 
is simply not realistic.
In environments where agents hold an amount 
of a divisible resource~(e.g., time, money, computational power),
which they can invest to earn utility,
it is natural for them to
 divide that resource in order to simultaneously
participate in a number of \emph{overlapping 
coalitions} \cite{shehory1998,dang2006,chalkiadakis2010,zick2012,zick2011,zick2014,mamakos2017},
to  maximize their profits.

As real-world environments exhibit a high level of uncertainty,
it is more natural than not to assume that agents do not have complete knowledge 
of the utility that can be yielded by every possible team of agents~\cite{suijs1999,kraus2003,chalkiadakis2004,ieong2008}.
Moreover, coalitional value is often determined by 
an underlying structure defined given
\emph{relations} among the members of the coalition.
These relations reflect the \emph{synergies}
among the coalition members.
It is natural to posit that agents do not know the exact synergies
at work in their coalitions.
Against this background, in our system the coalitional value depends on the amount
of resources the agents invest, 
and, crucially, the explicit relations among
coalition members.
As such, we build on the idea of 
\emph{marginal contribution nets}~(MC-nets)~\cite{ieong2005}
and introduce \emph{Relational Rules (RRs)},
a representation scheme for cooperative games 
with \emph{overlapping} coalitions.
The RR scheme allows for
the concise representation of the 
synergies-dependent coalition value. 

Now, an agent can make an observation of the utility that can be earned 
by the resource offerings of the members of a coalition, but it is a much more
complex task to determine her relations with subsets of agents of that coalition.
Probabilistic topic modeling (PTM)~\cite{blei2012} is a form of
\emph{unsupervised learning} 
which is particularly suitable for
 unravelling information from massive sets of documents.
Probabilistic topic models infer the probability 
with which each word of a given ``vocabulary'' is part of a
 \emph{topic}.
Intuitively, the words that have high probability in a topic,
 are very likely to appear
together in a document that refers to this topic with high probability.
Therefore, a topic, which is essentially a probability distribution
of the words of a given vocabulary, reveals 
the underlying
\emph{hidden structure}.
One of the most popular PTM algorithms~\cite{blei2012} is
 \emph{online Latent Dirichlet Allocation} (online LDA)~\cite{hoffman2010},
which, as its name indicates, is a an online version of
 the well-known
\emph{Latent Dirichlet Allocation} (LDA)~\cite{blei2003}
 algorithm.
LDA is  a generative probabilistic model for sets of discrete data,
while
online LDA can handle documents that arrive in streams,
 enabling the continuous evolution of the topics.

The method we develop employs online LDA to allow agents
to learn how well they can cooperate with others.
In our setting, agents \emph{repeatedly} form overlapping
coalitions, as the game takes place over a number of
\emph{iterations}.
Thus, we utilize a simple, yet appropriate, protocol, 
under which in each iteration an agent 
is (randomly) selected in order to propose 
(potentially) overlapping coalitions.
Agents that use our method take decisions on which coalitions 
to join
by exploiting the topics of the model that they have learned via employing online LDA: by
 interpreting formed coalitions as documents,
represented given an appropriate vocabulary,
 agents are able to use online LDA to update beliefs regarding the hidden
 collaboration structure---and thus implicitly
 learn rewarding synergies with others~(synergies which are in our experiments
 described by RRs).
Moreover, agents are able to gain knowledge
regarding coalitions that are costly, and should
thus be avoided.
Hence, agents can, over time, 
pick partners with which to cooperate
effectively. 
We have evaluated our approach against
two reinforcement learning (RL)
algorithms we developed for this setting,
and which serve as baselines.
Our algorithm vastly outperforms the baselines,
implying a high degree of accuracy in 
the beliefs of the agents,
 and a high quality
of  agent decisions.

To the best of our knowledge, the recent work of \cite{mamakos2017}
is the only one that has so far approached
\emph{overlapping} coalition formation {\em under uncertainty},
but it is concerned with the class of Threshold Task Games~\cite{chalkiadakis2010},
 which greatly differs to the more general setting we study here.
Moreover, ours is the first paper that
employs probabilistic topic modeling for multiagent learning:
existing literature on multiagent
learning~\cite{fudenberg1998,tuyls2012}, in both non-cooperative~\cite{littman1994,hu1998,hu2003} and cooperative~\cite{kraus2003,kraus2004,chalkiadakis2004,balcan2015} game settings,
is largely preoccupied with the study
of RL, PAC learning, or simple belief updating algorithms. As such, this paper introduces an entirely novel paradigm for (decentralized) learning employed by rational autonomous decision makers in multiagent settings.


\section{Background and Related Work}

In this section, we provide an overview of previous work on overlapping coalition formation, multiagent learning
and agent decision-making under uncertainty.
Furthermore, we offer the necessary background on Probabilistic Topic Modeling,
and in particular (online) Latent Dirichlet Allocation---which is employed in our proposed agent-learning method.


\subsection{Overlapping Coalition Formation}

Overlapping coalition formation was initially studied in~\cite{shehory1998}, 
which provided an approximate solution 
to the corresponding optimal coalition structure generation
problem~\cite{rahwan2015}, in a
setting where the costs of the coalitions and 
the capabilities of the agents are globally announced.
The method proposed in~\cite{shehory1998}
employs concepts from combinatorics and
approximation algorithms. 
Though related, our approach differs in that it
 is \emph{decentralized}, since the
 (overlapping) coalitions
 are formed by the agents themselves, and  are
 \emph{not} provided for the agents by an algorithm.
The subsequent work of~\cite{dang2006} presented an application of
overlapping coalitions in sensor networks.
An approximate greedy algorithm
with worst-case guarantees is introduced,
and
constitutes a real-world example of employing overlapping coalitions.
However, in that work, 
the agents do not form coalitions acting in a completely autonomous manner,
since they are, at a step of the algorithm, hardwired to agree on taking a specific action~(regarding
the choice of the members of the coalition).

As illustrated by the work
 of~\cite{chalkiadakis2010} which formally introduced cooperative games with overlapping coalitions (or OCF games),
whenever an agent can be part of a number of coalitions simultaneously, coalition structures
are much more complex than in non-OCF games---and so is the concept of deviation.
We provide a further discussion on the richness of OCF games
later in Section~\ref{sec:discussion} of this paper.
Furthermore, 
in~\cite{chalkiadakis2010} an expressive class of OCF games,
\emph{threshold task games} (TTGs),
is also presented.
In TTGs,
a coalition achieves a task and earns utility $u$ if its members manage to collect
a number of resource units which exceeds a threshold $T > 0$.

TTGs provide the framework of study for the work of~\cite{mamakos2017}. 
In that work, probability bounds for the resources contributed by members of overlapping coalitions are computed,
and subsequently exploited to form (overlapping) coalitions
that are deemed, with some probabilistic confidence,
capable of carrying over assigned tasks since they are believed
to possess resources exceeding the required threshold.
The paper uses Bayesian updating to update agent beliefs regarding partners' resources 
following coalition formation and task
execution, but no actual machine learning technique is used in that work.

In a series of works~\cite{zick2012,zick2011,zick2014} following~\cite{chalkiadakis2010}, Zick and colleagues study {\em stability}~\cite{chalkiadakis2011} with respect to the
behaviour of non-deviating players towards deviators in OCF settings.
Several variants of the core are developed, and the approach is based on
the notion of \emph{arbitration functions}, which define the payoff of the deviators
according to the attitude of the non-deviators.

A class of games highly related to cooperative games with overlapping coalitions,
is that of \emph{fuzzy coalitional games}~\cite{aubin1981}.
In a fuzzy game, an agent can be part of a coalition at various levels.
Thus, the coalitional value of $C \subseteq N$ is defined by the level at which 
the agents have joined $C$.
There is a number of differences between overlapping coalition formation and fuzzy games,
the biggest one being that in fuzzy games the core is the only acceptable outcome.
Finally, coalition structure generation with overlapping coalitions is studied in~\cite{zhang2010},
where a metaheuristic is developed, based on particle swarm optimization~\cite{eberhart1995}.


\subsection{Uncertainty and Learning}

Stochasticity in the value of payoffs in non-overlapping cooperative 
games has been studied in~\cite{suijs1999}, in a setting where agents 
have different preferences over a set of random variables.
The focus of that study is on core-stability.
Bayesian coalitional games are introduced in~\cite{ieong2008}, where
suitable variations of the core are also defined.
In \cite{kraus2003, kraus2004} agents have incomplete information regarding the costs
that the other agents incur by performing a task within a coalition,
while the formation of the coalitions takes place through
information-revealing negotiations and the conduction of auctions.
The formation of overlapping coalitions
 is not allowed in~\cite{kraus2003, kraus2004}.

One of the very early attempts in approaching learning in
 cooperative settings was presented in~\cite{claus1998}, 
 where the dynamics of a set of RL algorithms~\cite{sutton1998} were studied.
A Bayesian approach to reinforcement learning for coalition formation is 
presented in~\cite{chalkiadakis2004},
along with the introduction of a variation of the core.
The more recent work of~\cite{balcan2015} explores a PAC (probably
approximately correct) model for obtaining theoretical
predictions for the value 
of coalitions that have not been observed in the past.
The links between evolutionary game theory and multiagent reinforcement learning
is the topic of study in~\cite{tuyls2007, kaisers2009}.

Multi-agent learning in non-cooperative games~\cite{fudenberg1998} has been studied for a longer time.
Much  of the early seminal work~\cite{littman1994,hu1998,hu2003} 
is interested in the study of Q-learning algorithms and 
their convergence to 
Nash equilibria~\cite{nash1951}.
In particular, the algorithm presented in~\cite{hu1998} 
is shown to converge to a Nash equilibrium if every state
and action has been visited infinitely often and the learning rate satisfies some conditions regarding
the values it takes over time.
Overall, the literature on multiagent
learning~\cite{tuyls2012}, in both cooperative and non-cooperative settings,
is largely concerned with the study of reinforcement learning algorithms.



\subsection{Probabilistic Topic Modeling}

Probabilistic topic models~(PTMs)
 consist of statistical methods that analyze words of
documents, in order to discover the topics (or themes) to which these refer to,
and the ways the topics  interconnect.
One hugely popular and successful PTM
 is the Latent Dirichlet Allocation (LDA) \cite{blei2003}.

\subsubsection{Latent Dirichlet Allocation}
We begin by defining basic terms, following~\cite{blei2003,blei2012}:

\begin{itemize}
\item
A \emph{word} is the basic unit of discrete data.
A vocabulary consists of words and is indexed by $\{1,2,\ldots,V\}$,
while it is fixed and has to be known to the LDA.
\item
A \emph{document} is a series of $L$ words, denoted by
$\bm{w} =  (w_1, w_2, \ldots, w_L)$,
where the $l^{th}$ word is denoted by $w_l$.
\item
A \emph{corpus} is a collection of $M$ documents, denoted by 
$D = \{\bm{w_1}, \bm{w_2}, \ldots, \bm{w_M}\}$.
\item
A \emph{topic} is a distribution over a vocabulary.
\end{itemize}

LDA is a Bayesian probabilistic model,
 the intuition behind it being that a document 
 is a mixture of topics.
For each document $\bm{w}$ in $D$, LDA assumes a generative process
where a random distribution over topics is chosen,
and for each word in $\bm{w}$ a topic is chosen from that topics' distribution, finally choosing a word from that topic.
Documents share the same set of
topics, but exhibit topics in different portions.

While LDA observes only series of words, its objective
is to discover the \emph{topic structure} which is 
\emph{hidden}.
It is thus assumed that the generative process
 includes latent variables.
The topics are $\beta_{1:K}$, where $K$ is their number;
each topic $\beta_k$ is a distribution over the vocabulary, where $k \in \{1,\ldots,K\}$;
and $\beta_{kw}$ is the probability of word $w$ in topic $k$.
For the $d^{th}$ document the topic proportion of topic $k$ is $\theta_{dk}$,
as $\theta_d$ is a distribution over the topics.
The topic assignments for the $d^{th}$ document are denoted by $z_d$,
with $z_{dl}$ being the topic assignment for the $l^{th}$ word 
of the $d^{th}$ document. 
Thus, $\beta, \theta$ and $z$ are the latent variables of the model,
while the only observed variable is $w$, where $w_{dl}$
is the $l^{th}$ word observed in the $d^{th}$ document.
Given the documents, the posterior of the topic structure is:
\begin{equation*}
p(\beta_{1:K}, \theta_{1:D}, z_{1:D} \mid w_{1:D}) = 
\frac{p(\beta_{1:K}, \theta_{1:D}, z_{1:D}, w_{1:D})}{p(w_{1:D})}
\end{equation*}
where the computation of $p(w_{1:D})$, the probability of
 seeing the given documents under any topic structure,
 is intractable~\cite{blei2012}.
Furthermore, LDA introduces priors,
 so that
$\beta_k \thicksim Dirichlet(\eta)$ and $\theta_d \thicksim Dirichlet(\alpha)$.

Though the exact computation of the posterior, and thus the topic
structure as a whole, cannot be efficiently computed, 
it can be approximated~\cite{blei2003}.
The two most prominent alternatives for this are  
Markov Chain Monte Carlo (MCMC)
sampling methods~\cite{jordan1998}
and variational inference~\cite{jordan1999}.

In variational inference for LDA,
the true posterior is approximated
by a simpler distribution $q$
that depends on parameters (matrices)
 $\phi_{1:D}$, $\gamma_{1:D}$ and $\lambda_{1:K}$, defined as follows:
\begin{gather*}
\phi_{dwk} \propto exp\{E_q[log\hspace{1mm}\theta_{dk}] + E_q[log\hspace{1mm}\beta_{kw}]\} \\
\gamma_{dk} = \alpha + \sum_w n_{dw}\phi_{dwk} \\ 
\lambda_{kw} = \eta + \sum_d n_{dw} \phi_{dwk}
\end{gather*}

The variable $n_{dw}$ is the number of times that word $w$ has been
observed in document $d$.
Parameters $\gamma_{1:D}$
and $\lambda_{1:K}$ are associated with $n_{dw}$, while $\phi_{dwk}$ denotes
 the probability~(under distribution $q$) that 
 the topic assignment of
 word $w$ in document $d$ is $k$~\cite{blei2003}.
The variational inference algorithm 
minimizes the {\em Kullback-Leibler divergence}
between the variational distribution and the true posterior.
This is achieved 
via iterating  
between assigning values
to document-level variables and updating topic-level variables.

\subsubsection{Online Latent Dirichlet Allocation}
In online LDA~\cite{hoffman2010}, documents can arrive in batches~(streams),
and the value of $\lambda_{1:K}$ is updated through analyzing
 each batch of documents.
The variable $\rho_t = (\tau_0 + t)^{-\kappa}$ controls the rate at which
the documents of batch $t$ impact the value of $\lambda_{1:K}$.
Furthermore, the algorithm (Alg. 1) requires an estimation, at least, of the
total number of documents $D$, in case this is not known in advance.
The values of $\alpha$ and $\eta$ can be assigned once and remain
fixed.
Essentially, the probability of word $w$ in topic $\beta_k$,
 can be estimated as $\beta_{kw} = \lambda_{kw} / \sum \lambda_k$.
 
\begin{algorithm}
\label{alg:1}
\SetAlgoNoEnd%

	Initialize $\lambda$ randomly \\
	\For{t = 1 to $\infty$}
	{
		$\rho_t = (\tau_0 + t)^{-\kappa}$ \\	
		E step : \\
		Initialize $\gamma_{tk}$ randomly\\
		\Repeat{$\frac{1}{K}\sum_k |$change in $\gamma_{tk}| < \epsilon$}
		{
				Set $\phi_{twk} \propto 
				exp\{E_q [log \hspace{0.6mm} \theta_{tk}] 
				+ E_q[log \hspace{0.5mm} \beta_{kw}]\}$ \\
				Set 
				$\gamma_{tk} = \alpha + \sum_w n_{tw} \phi_{twk}$ \\
		}	
		M step : \\
		Compute $\tilde{\lambda}_{kw} = \eta + D n_{tw} \phi_{twk}$ \\
		Set $\lambda = (1 - \rho_t) \lambda + \rho_t \tilde{\lambda}$
	}
\caption{Online Variational Inference for LDA~\protect\cite{hoffman2010}.}
\end{algorithm}
\section{Relational Rules}

Agents have to form coalitions under what we term
\emph{structural uncertainty}.
This notion describes the uncertainty agents face regarding the value
of \emph{synergies} among them.
Such synergies are, in a non-overlapping setting,
concisely described by \emph{marginal contribution nets}~(MC-nets).
In MC-nets, coalitional games are represented by a set of rules
of the  form 
$Pattern \rightarrow value$,
where $Pattern$ is a conjunction of literals~(representing the participation
or absence of agents),
and applies to coalition $C$ if $C$ satisfies $Pattern$,
with $value \in \real$ being added to the coalitional value of $C$.

We now extend MC-nets to overlapping environments
by introducing
\emph{Relational Rules}~(RR), with the following form:
\begin{equation*}
A
\rightarrow
\frac
{\sum_{i \in A} \pi_{i, C}}
{|A|} \cdot value
\end{equation*}
where
$A \subseteq N$~(with $N = \{1, \ldots, n\}$ being the set of agents),
 $value \in \real$;
$C \subseteq N$
 is a coalition such that 
$A \subseteq C$;
$\pi_{i,C}$ is the portion of her resource that $i$ has
invested in coalition $C$:
i.e., $\pi_{i,C} = r_{i,C} / r_i$, where $r_i$ is the total resource
quantity~(continuous or discrete) that $i$ holds and $r_{i,C}$ 
is the amount she has
invested in $C$.
Therefore, $\pi_{i,C} > 0$, since $i \in C$~($r_{i,C}$ = $0$ essentially
means that $i \notin C$), and
$\pi_{i,C} \leq 1$, since $i$ can offer to $C$ at most $r_i$.

A rule applies to coalition $C$ if and only if 
$A \subseteq C$,
and in that case utility
$
\frac
{\sum_{i \in A} \pi_{i, C}}
{|A|} \cdot value
$
is added to the coalitional value of $C$.
Note that it is {\em not} required that an agent's
total resource quantity
$r_i$ has to be communicated to $C$'s other members, since a rule is applied by the environment.
In non-overlapping games, RRs reduce to MC-nets rules
 without negative literals,
as it then holds that
$
\frac
{\sum_{i \in A} \pi_{i, C}}
{|A|}
= 1
$.
\begin{example}
Assume that $N = \{1, 2, 3\}$,
$r_1 = 10, r_2 = 8$, $r_3 = 6$,
and the Relational Rules of the game are:

\begin{flalign*}
{ \emph{(1)}}: & \hspace{2mm}  \{1\}
\rightarrow
\pi_{1,C} \cdot 100 
\\
{ \emph{(2)}}: & \hspace{2mm}
 \{1, 2\}
\rightarrow
\frac{\pi_{1,C} + \pi_{2,C}}{2} \cdot (-50)
\\
  { \emph{(3)}}: & \hspace{2mm}
\{2, 3\}
\rightarrow
\frac{\pi_{2,C} + \pi_{3,C}}{2} \cdot 50
\end{flalign*}

Let coalition $C_1$ = $\{1, 2\}$ form,
with $r_{1, C_1}$ = $5$~($\pi_{1,C_1}$ = $0.5)$ 
and $r_{2, C_1}$ = $8$~($\pi_{2, C_1}$ = $1)$.
The value of $u_{C_1}$ will be determined by rules
\emph{(1)} and \emph{(2)},
 since rule $(3)$ does not apply, as agent $3 \notin C_1$.
Applying rule \emph{(1)} to $C_1$ will result in value 
$\pi_{1,C} \cdot 100 = 0.5 \cdot 100 =  50$ 
and applying rule \emph{(2)} to $C_1$ will result in value
$\frac{\pi_{1, C_1} + \pi_{2, C_1}}{2} \cdot (-50) =
\frac{0.5 + 1}{2} \cdot (-50) = -37.5$.
Thus, $u_{C_1} = 50 - 37.5 = 12.5$.

%
\end{example}

In our setting, the value of a coalition 
is determined through RRs,
but agents \emph{do not know the RRs in effect},
 and hence cannot
determine the value of a coalition with certainty.
Thus, agents do not know how
well they can do with others, 
and  cannot determine their relations
just by an observation of a coalitional value.
However, in Section 4 we show
how PTMs can 
be exploited so that agents 
 learn the underlying RR-described
collaboration structure.

\section{Learning by Interpreting Coalitions as Documents}


In this section, we present how agents can employ online LDA
in order to effectively learn the underlying collaboration structure.
We let each agent maintain and train
her own online LDA model.
Thus, there are $n$ such models in the system.
The agents' formation decision-making process (Section~\ref{sec:formation}) employs the learned topics.

For each (possibly overlapping) coalition $C \subseteq N$ formed,\footnote{Note that to improve readability we use the set notation ``$C \subseteq N$'' to refer to coalitions that can in reality be overlapping: these are in fact vectors of the resource quantities that each agent contributes to this coalition, i.e. a coalition is a $\vecr = \langle r_1 , \cdot , r_n\rangle$ vector~\cite{chalkiadakis2010}.} 
$i \in C$, 
$i$ observes the earned utility $u_C$.
The contribution $r_{i,C}$ 
of agent $i$ to coalition $C$
 is known to each other agent $j \in C \setminus i$,
 once $C$ is successfully formed.
However, in order to supply that information to her 
online LDA model,
an agent must maintain a vocabulary.
We define the vocabulary of an agent to include $n$ words,
one for each agent~(including herself), indicating their contribution,
plus two words for the utility, one representing gain and the other
representing loss, since the value earned from a coalition
can be either positive or negative.
Therefore, the vocabulary of an agent consists of $n+2$ words.
Assuming a game that proceeds in rounds,
in round $t$ agent $i$ interprets 
the coalitional configuration regarding $C, i \in C$, 
as a document by ``writing'' in the document the word that indicates the
contribution of agent $j \in C$ \hspace{0.7mm} $r_{j,C}$ times---where 
$r_{j,C} \in \natur_+$ is
the contribution of $j$ to $C$.
The restriction of the resource contributions of agents to
positive~($r_{j,C} = 0 \Rightarrow j \notin C$) natural numbers
is thus necessary 
when LDA is used,
since a word can only appear in a document
 a discrete number of times.
Thus, agent $i$ ``writes'' in the document,
that corresponds to $C$, either the word that indicates gain
or the one that indicates loss as many times as the absolute value of the utility earned by the coalition is.\footnote{The number 
of times that the word for utility
is written may require scaling when its domain ranges from very low to very high values.
}
Since words are discrete data, $u_C$ cannot be real-valued; so,
we let the actual value earned by $C$ 
be $\floor{u_C}$, instead of the  $u_C$ computed by 
the application of the RRs related to $C$. The number of documents that an agent passes in an
 iteration~(round) to her online LDA is equal to the number of coalitions
 that she is member of.
\begin{example}
\label{ex-2}
Let an agent's vocabulary include the words
``ag1'', ``ag2'', ``gain'' and ``loss'',
corresponding respectively to
 agents' $1$ and $2$ contribution and the positive and negative utility.
Thus, for coalition $C$ = $\{1,2\}$
where $r_{1,C}$ = $3$, $r_{2,C}$ = $1$, and $u_C$ = $-3$, 
each agent $i \in C$ forms the document:
\begin{equation*}
\textbf{w} = (\text{``ag1'', ``ag1'', ``ag1'', ``ag2'', ``loss'', ``loss'', ``loss''})
\end{equation*}
\end{example}

\begin{example}
Following Example \ref{ex-2}, let also agent $3$ participate in the game,
and so an agent's vocabulary includes the words
``ag1'', ``ag2'', ``ag3'', ``gain'' and ``loss'',
with the corresponding meaning of each being as defined in Example \ref{ex-2}.
Now, for coalition $C$ = $\{1,2, 3\}$
where $r_{1,C}$ = $1$, $r_{2,C}$ = $1$, $r_{3,C} = 2$, and $u_C$ = $4$, 
each agent $i \in C$ forms the document:
\begin{equation*}
\textbf{w} = (\text{``ag1'', ``ag2'', ``ag3'', ``ag3'', ``gain'', ``gain'',
 ``gain'', ``gain''})
\end{equation*}
\end{example}

Since LDA is a ``bag-of-words'' model, 
the order of the words in the document 
does not matter. 
The batch of documents the online LDA model of agent $i$ 
is supplied with at iteration $t$,
consists of the interpreted-as-documents coalitions
 $i$ has joined at $t$.
The intuition behind the notion of a topic is that the words  
that appear in it
with high probability are very likely to 
appear together in a document
that exhibits this topic with high probability.
Thus, \emph{the probability with which the word corresponding 
to an agent's 
contribution appears in a topic, is correlated
with the amount of her contribution}.
Therefore, the meaning of a topic identified by agent $i$,
 is that $i$ has observed 
 in many documents certain agents who contributed
a lot, and some that contributed less; and
this configuration \emph{results to gain or loss
with the corresponding probabilities}.

\begin{figure}[H]
	\centering
    \subfloat[A ``profitable'' learned topic.]			  
    {{  
    \includegraphics[width=10cm]{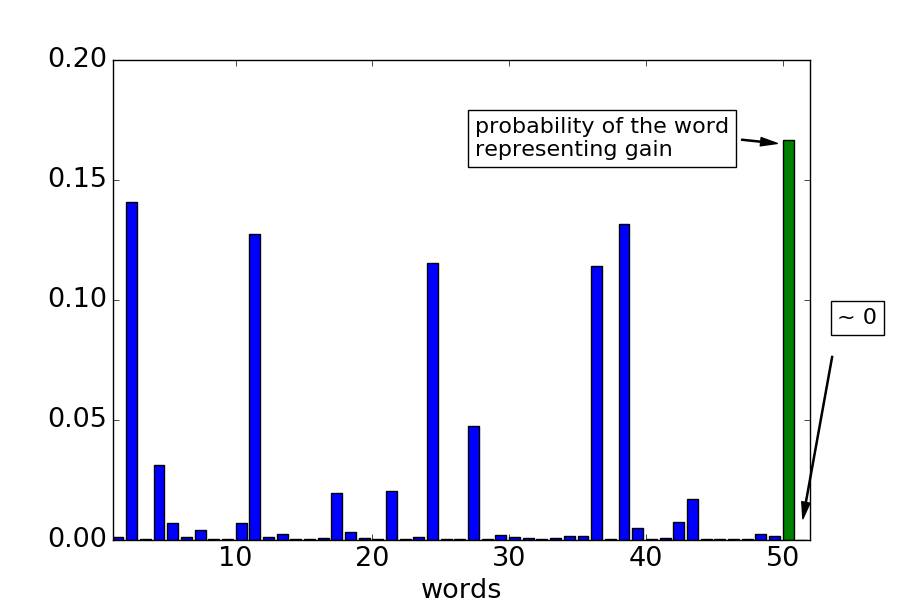} }}%
    \\
    \subfloat[A ``non-profitable'' learned topic.]
    {{ 
    \includegraphics[width=10cm]{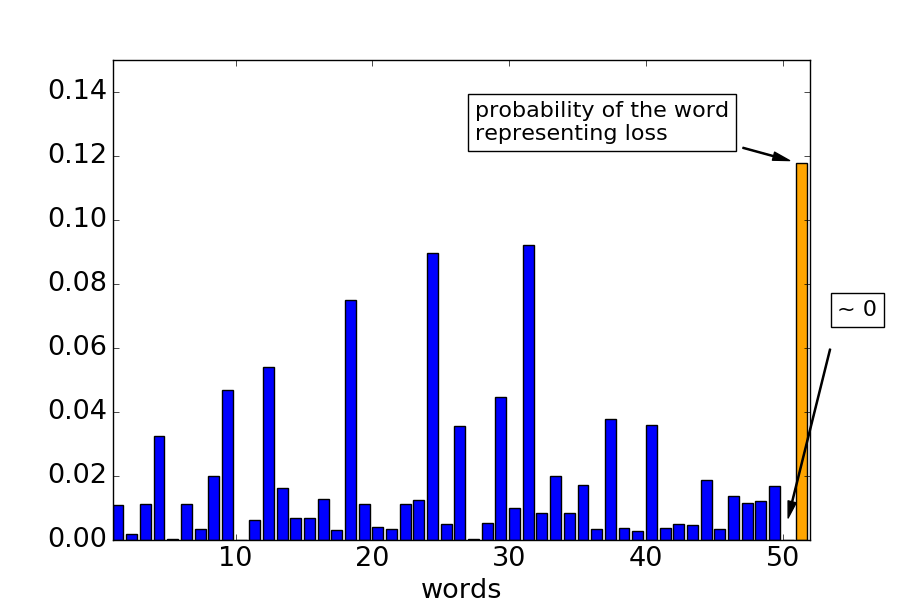} }}%
    \caption{Typical topics, as formed by a randomly selected agent at 
    the end of a random
    iteration in an experiment, where
   an agent's vocabulary consists of $52$ words ($n = 50$).
 The two last words in a topic indicate the probability 
 of gain and loss respectively, while the rest correspond to
    agents' contribution.
In (a), the ``profitable'' learned topic, the word for loss appears 
with near-zero probability; in (b), the ``non-profitable'' topic, 
the word for gain has near-zero probability.}
    \label{fig:topics}%
\end{figure}

Thus, the topic in Fig.~\ref{fig:topics}(a) implies that if $i$ joins
a coalition with
 the agents that appear in the topic with high probability, 
then that coalition would be profitable.
On the other hand, 
the topic in Fig.~\ref{fig:topics}(b) implies
that forming a coalition with the
agents that appear in it with high probability
would result in loss.
Note that learning a topic's profitability corresponds to 
acquiring information on the RRs associated
with that topic.
However, these RRs are not explicitly learned; what is learned
is the \emph{underlying collaboration structure}~(which might, in the general case, be
generated by means {\em other} than RRs).
It is natural to expect that agents who appear with (relatively) high probability
in a topic which has been associated with loss,
like the one in Fig.~\ref{fig:topics}(b),
will not appear~(as a group) with high probability in a topic that
has been associated with gain, like the one 
in Fig.~\ref{fig:topics}(a).
Such occurrence would reflect that an agent's beliefs indicate
that cooperation with a group of agents is~(paradoxically)
both beneficial and harmful.
Furthermore, as an agent observes documents that always include the
word that corresponds to her own contribution,
it is expected that her corresponding word will have a non-trivial
probability in her topics.

\section{Taking Formation Decisions}
\label{sec:formation}

We now present \texttt{OVERPRO}, a method for agent decision-making
in 
iterated
OVERlapping coalition formation games, 
via PRObabilistic topic modeling~(here, online LDA).

\subsection{A Repeated OCF Protocol}
The protocol of our game operates in $I$ iterations~(rounds).
At the beginning of an iteration one agent is randomly selected,
from the set of agents $N = \{1, \ldots, n\}$, as the proposer.
Then, this agent proposes a number of (overlapping) coalitions, 
where for each such coalition she offers 
an integer quantity of her resource
and asks for a (possibly different) resource quantity from each agent
of each coalition. 
Therefore, proposer $i \in N$ is asked to pass a 
list of tuples of the form
$\langle demands_C, r_{i,C} \rangle$, 
where $demands_C$ is an $n$-dimensional vector
whose $j^{th}$ entry denotes the (integer) resource quantity
that the proposer asks from agent $j$ for joining $C$,
and $r_{i,C} \in \natur^+$ denotes the amount of resource
that $i$ offers to coalition $C$. 
Naturally, if the proposer does not ask from $j$
to participate in $C$, then the $j^{th}$ entry of $demands_C$
is $0$.
By limiting the agents' resource investments to discrete quantities
we disallow the formation of an infinite number
of coalitions.
Then, every agent $j \in N \setminus i$ 
is a responder and
gets  informed of the proposals in which she is involved,
while she has to respond to each such proposal by either
accepting (and thus offering the requested resource quantity)
or rejecting it.
A (possibly overlapping) coalition $C$ forms if and only if 
all involved agents accept to participate in it.

At the end of each round, all coalitions are dissolved and the resources of the agents are replenished.
This removes the need for long-term strategic reasoning by the agents---and thus removes unnecessary distractions from the study of the effectiveness of the method used for learning the collaboration structure (which is what we focus on in this paper).

The utility $u_{i,C}$ that $i$ $\in C$ earns 
from coalition $C$ 
is \emph{proportional} to her $r_{i,C}$ contribution,
i.e., $u_{i,C} = u_C \cdot r_{i,C}/\sum_{j \in C} r_{j,C}$,
where $u_C$ is the total utility earned by the coalition.\footnote{The use of more elaborate reward allocation methods is interesting future work.}
An agent receives information regarding partners' contributions to, and the total coalitional utility of,
her own formed coalitions only.

\subsection{The OVERPRO Method}

The main idea behind \texttt{OVERPRO} is that an agent exploits her learned LDA topic model
for profitable coalition formation.
Specifically, by considering as ``profitable''~(``non-profitable'')
 the topics in which
the probability of gain~(loss) is 
higher than 
the probability of loss~(gain),
the agent can identify  coalitions that
will potentially result in gain (loss). 
Now, it might be that not all of the topics are
\emph{significant},
in the sense that they are not clearly profitable or harmful.
This is because 
  some of them
might not be well formed~(especially at the early iterations of the game).
We define a topic to be \emph{significant} if the absolute
value of the difference between the probability of
the word representing gain and the probability of the word
representing loss is greater than $\epsilon$.
Therefore, the \emph{significant topics} of agent $i$ are:
\begin{equation*}
ST^i \leftarrow \{k: |\beta^i_{k,'gain'} - \beta^i_{k,'loss'}| >
\epsilon\}
\end{equation*}

\noindent where $\beta^i_k$ is the $k^{th}$ (out of the $K$) topic of agent $i$.
Furtheremore, we define \emph{Good}~(profitable)
 topics as the ones
 in which the probability of the word
 representing
 gain is greater than that of the word 
 representing loss, and are significant.
\emph{Bad} topics are defined analogously.
Formally:
\begin{gather*}
Good^i \leftarrow \{k: \beta^i_{k,'gain'} > \beta^i_{k,'loss'} \land
k \in ST^i\} \\
Bad^i \leftarrow \{k: \beta^i_{k,'gain'} < \beta^i_{k,'loss'} \land
k \in ST^i\}.
\end{gather*}

How much probability should an agent 
appear with in a topic in order for the \emph{agent} to be considered significant for that topic?
For instance, in Fig.~\ref{fig:topics}(a) not all agents
appear in the profitable topic with similar probability values.
Note that, due to initialization of Dirichlet distributions,
 each word appears in a topic with positive probability, no matter
how small.
We define the \emph{significant agents} of topic
$k$ of agent $i$, denoted as $SA^i_k$,
 as those whose corresponding words in topic $k$
have probability higher than the mean value $\mu$
of the probabilities of the words corresponding
to agents, plus the standard deviation $\sigma$ of those.
Formally:
\begin{gather*}
SA^i_k \leftarrow
\{j: \beta^i_{k, j} > \mu(\beta^i_{k, N}) + \sigma(\beta^i_{k, N})
\land j \in N \setminus i \}
\end{gather*}
Given the above,
 the approach of \texttt{OVERPRO} is that a proposer $i$
proposes one coalition for every topic $k \in Good^i$
~(proposing just the one coalition corresponding to the most profitable topic
is risky, since a single rejection would mean formation failure),
with the proposed members of the coalition stemming from topic $k$ being $SA^i_k$.
Then, the resource
quantity $r_{i,C}$ offered to $C$ by $i$ 
is proportional to 
$\beta^i_{k,i} \cdot ( \beta^i_{k, 'gain'} - \beta^i_{k, 'loss'} )$,
where $k$ is the topic in which coalition $C$ was identifibed as profitable.
Thus, a proposer's contribution to a coalition is affected by both
his effect on the corresponding topic, and the profitability of this
topic.
The proposer $i$ asks from agent $j \in C$ to offer to $C$
 the quantity $r_{j,C} = r_{i,C} \cdot
\beta^i_{k, j} / \beta^i_{k,i}$, 
since $i$ assumes that others will respond according
to the topics that $i$ has observed.


Now, an agent often faces the
dilemma to either exploit her best-so-far action, or
explore different options~\cite{sutton1998}.
We deal with this issue by allowing agent $i$ to do both at
the same time, since $r_i$ is divisible. 
Specifically, at iteration $t$
proposer $i$ dedicates $z_t \in (0, 1)$
of $r_i$ in exploring and $1-z_t$ in exploiting~(with the exploitation part defined as described above).
Then, an agent performs exploration by 
proposing to $\floor{r_i \cdot z_t}$ random coalitions, 
offering to each~(and asking from each one's
participating agents) the minimum possible resource quantity $(=1)$.

In each iteration, responders
receive the proposals in which they are involved, 
and decide, for each proposal in turn
whether to accept it~(invest the
resource quantity) or reject it~(offering nothing).
\texttt{OVERPRO} employs a parameter $c \in (0,1)$,
so that an agent rejects a proposed coalition $C$ 
if she identifies a non-profitable~(bad) topic 
in which at least $c$ of the agents 
in $C$ are significant.
The intuition behind the employment of parameter $c$
is that it suffices to observe a certain percentage of
agents of a proposed coalition in a ``non-profitable'' topic
in order to reject it.
Parameter $c$ can have different values at different iterations,
so we refer to its value at iteration $t$ as $c_t$.
As agents make more observations, and thus become more confident about their beliefs over time,
they gradually become more strict about who they cooperate with, and thus the value
of $c_t$ decreases with $t$.
If a proposal to form coalition $C$ is not rejected,
 then it is checked whether
there is a profitable~(good) topic in which
at least $(1 - c_t)$ of the agents
in $C$ are significant.
Since responder $j$ has to split her resources among the proposals
she has received,
a proposal associated with a profitable topic
like the one described above,
is an item of a \emph{KNAPSACK} problem that $j$
has to solve 
 where the value of the item is 
$r_{j, C} / \sum_{m \in C} r_{m,C} 
\cdot (\beta^j_{k, 'gain'} - \beta^j_{k,'loss'})$:
this stands for the profit portion of $j$, multiplied by the
profitability of the corresponding topic.
The weight of the item is the requested quantity $r_{j, C}$,
and the constraint is that responder $j$
cannot invest more than $r_j$ in total.
Thus, responder $j$ accepts the coalitions which correspond 
to the items given by solving the \emph{KNAPSACK} problem,
while she rejects the rest.
Despite that \emph{KNAPSACK} is an \emph{NP-Hard} problem,
the dynamic programming pseudopolynomial algorithm~\cite{knapsackbook},
which we used in our experiments,
often admits not excessive running times, while
an alternative is to use an \emph{FPTAS}.
After the decisions regarding the coalitions identified by profitable topics
are made,
the responder replies positively~(if there is sufficient resource quantity)
 to an offer which 
has been neither accepted nor rejected~(no relevant information found)
 if either the requested quantity is $1$, or, if not, with probability
$c_t$~(in an exploratory sense).

The training of LDA, and thus consequently that of online LDA, takes polynomial
time in the number of documents and topics~\cite{blei2003}.
Despite the fact that the number of documents depends on the resource quantities
of the agents, which are numeric values and thus imply pseudopolynomial
complexity, in practice the number of documents formed
 is far from the worst case, which can be attested by our experimental results.
The ``exploitation'' part of \texttt{OVERPRO} takes polynomial time
in the number of topics and agents,
while the exploration part, which is independent of \texttt{OVERPRO}
and can be replaced by one's choosing, takes pseudopolynomial time,
as it linearly depends on the proposer's resource quantity.

Furthermore, topics convergence is guaranteed~\cite{hoffman2010} if 
$\kappa \in (0.5, 1]$ (online
LDA parameter). Therefore, assigning such a value to $\kappa$ and letting the 
resource portion dedicated by an agent for exploration decrease
over time,
we have as a corollary that the actions of the agents employing 
\texttt{OVERPRPO} converge.

\section{Reinforcement Learning for OCF}

To the best of our knowledge, this is the first work 
on~(decentralized) multiagent learning for \emph{overlapping}
 coalition formation under uncertainty not restricted in the context
 of Threshold Task Games~\cite{mamakos2017},
  and thus
there is no algorithm to use as a means
for comparison.
To this end,
 we have developed a
 a Greedy top-k algorithm
 and a Q-learning style~\cite{watkins1992,claus1998} one,
 and use these as baselines.

\subsection{Greedy top-k algorithm}

An agent that uses our Greedy top-k algorithm
 maintains the \emph{k} most profitable
coalitions she has observed, along with their values, 
and the resources offered by participating agents.
A proposer makes \emph{k} proposals, one for each
of the top-k coalitions. 
The resource $r_{i,C}$ offered to coalition $C$ by proposer $i$
is proportional
to the
amount $i$ previously offered to $C$
and $C$'s corresponding value
 derived by applying the \emph{softmax function}~\cite{sutton1998} 
over the maintained values~(observed utilities) of the top-k coalitions;
while $i$ asks from agent $j \in C$ an amount equal to
 $r_{i,C}$, multiplied by the ratio of their previous offerings~(with $j$'s 
 being on the numerator and $i$'s on the denominator)
 and a random value in $U(0.9, 1.1)$.
The approach to the exploitation-vs-exploration problem is 
exactly the same as in \texttt{OVERPRO}, and thus
the $z_t$ parameter is also employed here.
A responder $j$ adds a proposal to the input of a 
\emph{KNAPSACK} problem
if at least $(1 - c_t)$ of the agents in the corresponding~(proposed)
coalition appear in one of the top-k coalitions.
The value of the \emph{KNAPSACK} item
is equal to the sum of the values of the coalitions in which the agents of 
the proposed coalition were identified, 
multiplied by $r_{j, C} / \sum_{m \in C} r_{m,C}$.
The weight of the corresponding item is, naturally,
the requested quantity $r_{j, C}$.
A responder accepts a proposal
which was not included in the input of the  \emph{KNAPSACK} instance,
and for which there is sufficient remaining resource,
if the requested quantity
 is $1$,
or else with probability $c_t$.

\subsection{Q-learning for OCF}

An agent that uses our Q-learning algorithm employs
two distinct kinds of Q-values.
The first one, denoted as $Q_a$, 
maintains agent-level values; while the second,
 denoted as $Q_s$, maintains coalition 
 size-level values.
Employing two different sets of Q-values is necessary, since 
the alternative of maintaining a Q-value for every possible coalition 
requires exponential space in the number of the agents~(rendering
the problem practically intractable in large settings).
Agent $i$ maintains for each agent $j \in C \setminus i$
a $Q^i_{a,j}$ value, 
and for each $h \in \{1, \ldots, n-1\}$ 
a $Q^i_{s,h}$ value;
keeping a $Q^i_{s,h}$ value for $h = n$ is redundant since
the decision-maker always includes herself in a coalition.
Furthermore, a learning rate $\delta_t \in (0,1)$ is employed~\cite{sutton1998},
as is common in Q-learning, where $t$ is the game iteration.
After $C \ni i $, is formed and coalitional value $u_C$ is observed, agent
$i$ updates her Q-values as follows:
\begin{gather*}
Q^i_{a, j} \leftarrow Q^i_{a, j} + \delta_t \cdot
(u_C  - Q^i_{a, j})
\hspace{2mm} \forall j \in C \setminus i \\
Q^i_{s, h} \leftarrow Q^i_{s, h} + \delta_t \cdot
(u_C - Q^i_{s, h})
\hspace{2mm} \text{where } h = |C| - 1
\end{gather*}

A proposer employing our Q-learning algorithm iteratively
selects some quantity of her resource, at random, to offer to a coalition,
until it is depleted.
Then, at each iteration, the size of the coalition to propose
 (excluding herself) is selected
using the \emph{softmax function}
 over the $Q^i_s$ values,
and afterwards the agents to include in the coalition
are selected using the \emph{softmax function} 
over the $Q^i_a$ values.
The proposer asks from each member in $C$ the same quantity
she has offered to $C$ multiplied by $U(0.9, 1.1)$.
Exploration is employed in the same way as in the other methods.


%
%
%

Responder $j$ has to solve a \emph{KNAPSACK} problem,
where a proposal regarding coalition $C$ is given as input only if
$\sum_{l \in C \setminus j} Q^j_{a,l}$ is positive,
 with its value being 
 $\sum_{l \in C \setminus j} Q^j_{a,l} \cdot$  $r_{j, C} / 
 \sum_{m \in C} r_{m,C}$
 and its weight $r_{j, C}$.
If $\sum_{l \in C \setminus j} Q^j_{a,l}$ is negative, $j$
 accepts (if she can afford it) joining $C$
  if the requested quantity is $1$, or else
 with probability $c_t$.


%
%
%
%
%

\section{Experimental Evaluation}

 We evaluated \texttt{OVERPRO}'s effectiveness and robustness in environments with 50 and 250 agents.
Agent resource quantities were generated from
$\{475, \ldots, 525\}$ uniformly at random.
The RRs were $500$
 for $n = 50$,
and $20,000$ for $n=250$,
where the value of each RR was generated from
$\mathcal{N} (0,100^2)$.
We added stochasticity in our setting, so that with probability $5\%$
the value of a coalition, as resulted by applying RRs, is multiplied by a
factor generated from $\mathcal{N} (0,5^2)$.
Every game ran for $I$ = 1000 iterations,
and thus, 
agent $i$ can observe at most 
 $r_i \cdot 1000$  documents.
Coding was in Python 3 and
online LDA was implemented as 
in~\cite{hoffman2010}.\footnote{https://github.com/blei-lab/onlineldavb}
The same exploration rate $z_t$ was set for
all methods,
 decreasing quadratically from $1$ to $10^{-3}$.
The value of $c_t$ decreases linearly, for every method as well,
from $1$ to $0.5$.
We tested 
\texttt{OVERPRO}, which requires
$K$ (number of topics),\footnote{In some
LDA implementations the value of $K$ is automatically derived~\cite{teh2012},
but we use the standard online LDA algorithm, which requires the value
of $K$ as a parameter.} $\tau_0$ and $\kappa$ (that
 determine the impact \mbox{$\rho = (\tau_0 +t)^{-\kappa}$}
 of a batch of documents on the topics),
 and Q-learning,
which requires $\delta_t$,
 for a number of different parameters;
 while for Greedy top-k we set values of \emph{k} equal
 to the number of topics $K$ for \texttt{OVERPRO}.
The value of $\epsilon$ used in \texttt{OVERPRO}
was equal to the vocabulary's length raised to $-1$, i.e., $(n+2)^{-1}$.
Experiments ran on 
a grid~(each execution instance ran sequentially) with 4GB RAM  2.6GHz computers.

\begin{table}
\caption
{
Results~(averages over $75$ runs) for
$50$ agents and
 different values of $\langle K ,\tau_0, \kappa \rangle$  for
\texttt{OVERPRO},
\emph{k} for Greedy top-k, and
and $\delta_t$ for Q-learning.
Participation 
and time
 are per agent per 
 iteration $t$~(there is a unique proposer in $t$).
 }
\center
\label{table:1}
\begin{tabular}{|c|c|c|c|}
\hline
 $n = 50$   & sw~($\cdot 10^3$) &  participation & 
    time~(sec) \\ \hline
$\langle 10, 100, 0.7  \rangle$ & 95.60  & 24.91 & 0.525 \\ \hline
$\langle 10, 200, 0.9  \rangle$ & 117.27  & 24.97 & 0.528 \\ \hline
$\langle 15, 100, 0.7  \rangle$ & 108.59  & 25.09 & 0.540 \\ \hline
$\langle 15, 200, 0.9  \rangle$ & 119.47  & 25.17 & 0.543 \\ \hline
$k = 10$                        & 34.54   & 37.70 & 0.363 \\ \hline
$k = 15$                        & 51.72   & 37.64 & 0.366 \\ \hline
$\delta_t = 0.95^t$             & 14.53    & 38.15 & 0.009 \\ \hline
$\delta_t = 0.99^t$             & 10.69    & 38.08 & 0.009 \\ \hline
\end{tabular}
\end{table}

In Table~\ref{table:1} 
we present 
for $n = 50$
the average: 
social welfare~(total utility) earned in a game~(sw);
number of coalitions in which an agent participates
in a round~(participation);
and game completion time 
per agent per iteration.

As observed in Table~\ref{table:1}, \texttt{OVERPRO} vastly outperforms
both Greedy top-k and Q-learning in terms of social welfare.
For the best set of parameters
of \texttt{OVERPRO}, $\langle K=15, \tau_0=200, \kappa=0.9 \rangle$,
 the average social welfare earned in a game was more than double
 of that earned when the best alternative was employed,
 which is Greedy top-$15$, since $119.47 / 51.72 = 2.3$.
Thus, we can conclude that a  stochasticity probability
even as low as $5\%$ can have a largely negative impact on Greedy top-k.
On the other hand, this demonstrates the robustness of \texttt{OVERPRO}.
Q-learning, for both values of $\delta_t$,
performed very poorly, as in both cases the social welfare was
not far above zero: $14.53$ for $\delta_t = 0.95$
and $10.69$ for $\delta_t = 0.99$.
This suggests deficiency in matching good agent-level Q-values
to coalition size-level ones, and thus unsuitability of 
Q-learning approaches, when Q-values for every coalition cannot be maintained.
For both $10$ and $15$ topics, the social welfare was better 
for $\langle \tau_0=200, \kappa=0.9 \rangle$ than for
$\langle \tau_0=100, \kappa=0.7 \rangle$.
Since $\tau_0$ and $\kappa$ determine the impact
that a batch of documents has on the formation of the topics,
 $\rho_t = (\tau_0 + t)^{-\kappa}$ can be interpreted as
a learning rate.
Now, higher values of $\tau_0$ and $\kappa$ result in smaller values
of $\rho_t$.
Therefore, it can be conjectured that lower learning rates are 
preferred over higher ones.

Despite the better social welfare performance
 of \texttt{OVERPRO} against any of the alternatives,
the average agent  participation per iteration
is lower when \texttt{OVERPRO} is employed,
as seen in Table~\ref{table:1}.
In particular, an agent using \texttt{OVERPRO} joins about $25$ coalitions
 per iteration, while one using
 either Greedy top-k or Q-learning joins about $38$.
By the end of a game an agent employing \texttt{OVERPRO} will have trained
her online LDA with more than $24.9$k documents, since one coalition
corresponds to one document, an agent participates in at least $24.91$
coalitions in a round~\big(value for $\langle K=10, \tau_0 = 100, \kappa=0.7 \rangle$\big), and $I$=$1000$.
Notice that the number of coalitions in which an agent participates
when \texttt{OVERPRO} is employed is much smaller 
than her resource quantity,
which is at least $475 >>25.17$ (= max \texttt{OVERPRO} participation). 
Thus, the number of documents is much smaller than the maximum possible,
which implies that the pseudopolynomial complexity related to the
number of documents does not have an actual impact. 
The time taken per agent per iteration is less 
than $0.55$ sec for \texttt{OVERPRO}
and $0.37$ sec for Greedy top-k,
while it is about two orders of magnitude lower
for Q-learning.
 
Now, one cannot draw accurate conclusions regarding the real power
of an agent decision-making algorithm relying solely on social welfare.
For instance, when more coalitions form,
this will likely have a positive impact on 
social welfare---but rational
agents aim to maximize their own utility.
Therefore, we define \emph{efficiency} as
the ratio of social welfare~(total utility)
 to total resource quantity invested
 by all agents in every coalition in a round.
This efficiency metric is natural, 
as
it takes the focus away from social welfare.



\begin{figure}[h]
	\centering
	\includegraphics[width=10cm]{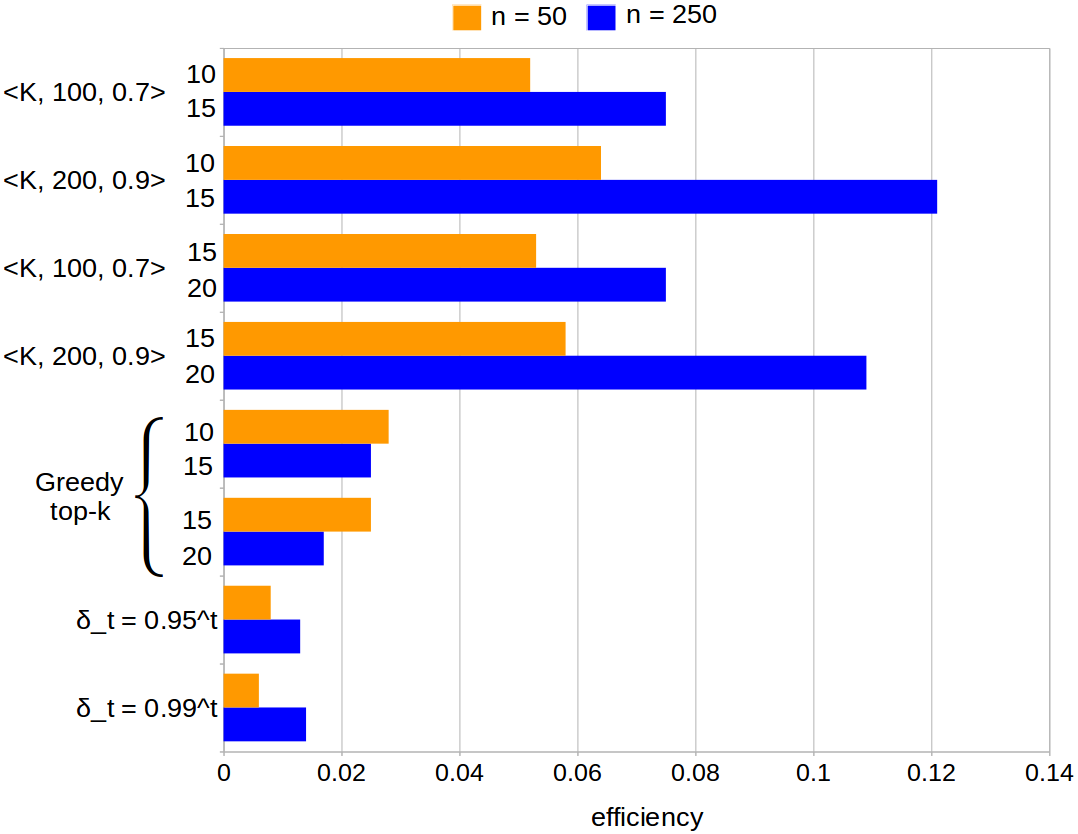}
	\caption{Average efficiency defined as the ratio of social welfare
	to total resource quantity invested by all agents.
	For \texttt{OVERPRO} the values of $K$~(number of topics),
	and respectively for Greedy top-$k$ the values of $k$,
	are denoted on the left of each bar.
	Results are averages over all rounds over multiple runs:
	$75$ runs for $n=50$, and $30$ for $n=250$.
	}
	\label{fig:uq}
\end{figure}

It can be observed in Fig.~\ref{fig:uq}
that \texttt{OVERPRO}, for $n = 50$~(orange bars),
outperforms both
Greedy top-k and Q-learning in terms of efficiency. 
In particular, 
the highest efficiency value, which appears for
$\langle K=10, \tau_0 = 200, \kappa=0.9 \rangle$,
is more than double of the best efficiency value
of the alternatives, observed for Greedy top-$10$~(the former 
being over $0.06$, and the latter lower
than $0.03$).
We can thus conclude that
agents employing \texttt{OVERPRO}
are more efficient in terms of earning 
 utility~(as a function of resources invested),
 as they focus more on coalitions identified 
 as profitable.

Now, as depicted by the blue bars in Fig.~\ref{fig:uq},
\texttt{OVERPRO} achieves even better efficiency 
for $n=250$ than for $n=50$,
and it thus appears to effectively exploit
the richer emerging collaboration structure.
Moreover, we observed through experimentation
that the number of topics $K$ should increase sublinearly
to $n$,
and we have thus used $15$ and $20$ as its values.
We set the values of $k$ for Greedy top-k accordingly.
We observe that for $\tau_0 = 200$ and $\kappa = 0.9$,
 for $K$ set to either $15$ or $20$,
the achieved efficiency is over $0.1$, thus vastly outperforming
the RL algorithms, whose efficiency is always lower than $0.03$;
and at the same time 
supporting
 our conjecture 
that lowering learning rates are associated with increased performance.
Agents employing Q-learning performed better
for $n = 250$ than for $n = 50$
in terms of efficiency, but their performance was still
very far below that of the ones using \texttt{OVERPRO}.
The efficiency 
 of Greedy top-k deteriorated for $n$ = 250~(dropping
  even below $0.02$ for $k = 20$),
as it fails to identify and exploit patterns
in the collaboration structure,
while it is more difficult for the method to identify profitable coalitions
in this larger setting.
Each agent using \texttt{OVERPRO} had trained her online LDA model
with more than $33.7$k documents
(coalitions) in total, at the end of a game.
The \texttt{OVERPRO} running time for $n=250$,
was $\sim1.15$ sec per agent per iteration.



\section{Discussion}
\label{sec:discussion}

One of the aims of this paper is to provide insights towards 
new research directions.
In particular, this work diverges from the standard RL/MDPs paradigm used for multiagent learning in games.
As such, we expect that it will raise intriguing questions, and bears the potential for the development of 
exciting new theory to be applied to challenging problems.
It is, for one, interesting to study the effect of adopting PTMs
for multiagent learning.
Taking a reverse point of view, it is of interest of examine
the effect that specific multiagent environments can have on PTM properties.
For instance, the convergence property of online LDA tranfers ``for free''
to \texttt{OVERPRO}, but in certain environments,
or under additional assumptions on the structure of the game, 
the convergence rate might be different.

Moreover, a somewhat ``orthogonal'' contribution in this paper was the 
introduction of the novel concept of \emph{Relational Rules (RRs)}, 
which consist
 a natural scheme for representing synergies in overlapping settings.
As such, pursuing their further study could lead
to new stability results for overlapping cooperative games.

Indeed, the concept of stability
in games with overlapping coalitions is quite elaborate and different than the disjoint coalitional ones,
since it is not just the membership of an agent in a coalition 
that matters, but the degree by which she participates in that; and since the number of different coalition structures cannot be enumerated
in such settings~\cite{chalkiadakis2010}. 
Additionally, in such settings, it is not just agents or coalitions that can deviate,
but entire coalition structures, since agents
 can withdraw just a portion
of their resource from the formed coalitions.
All these render the study of stability challenging in OCF domains.
Though we did not pursue the study of OCF stability in this paper, 
it would be interesting to define and study stability concepts that take into account agent preferences
regarding collaboration structures learned using our method.

\section{Conclusions and Future Work}

We have presented a novel approach for
multiagent learning in cooperative game environments,
 where probabilistic topic modeling
is exploited.
Furthermore, this is the first work
to tackle overlapping coalition formation
under uncertainty,
where the uncertainty is on the relations
entailing synergies among the agents.
To this end, we first proposed
Relational Rules, a
representation scheme 
which extends MC-nets to 
 cooperative games with overlapping coalitions;
and then showed how to use online LDA to implicitly learn
the agents' synergies described by~(unknown) RRs.
Simulations confirm the method's effectiveness.
 
As immediate future work, we intend to test these ideas in non-transferable 
utility settings.
Moreover, we would like to apply our method to non-cooperative environments.
Naturally, this would require 
adjustments to  
the ``vocabulary'' used in ``documents'' representing coalitions.
Finally, we intend to apply alternative PTM algorithms, 
to this or different game theoretic settings.

 



\balance
\bibliography{ocfptm}
\bibliographystyle{plain}

\end{document}